\documentclass[aps,twocolumn,showpacs]{revtex4}
\usepackage{graphicx}   
\usepackage{latexsym}   

\newcommand{\bold}[1]{\mbox{\boldmath $#1$}} 
\newcommand{\MeV}{{\rm MeV}} 
\newcommand{\p}{\bold{p}}
\newcommand{\bfv}{\bold{v}}
\newcommand{\y}{{\sf y}}

\begin{document}

\title{Signals of spinodal phase decomposition 
in high-energy nuclear collisions}

\author{J\o rgen Randrup}

\affiliation{Nuclear Science Division, 
Lawrence Berkeley National Laboratory, Berkeley, California 94720, USA}

\date{\today}

\begin{abstract}
High-energy nuclear collisions produce quark-gluon plasmas
that expand and hadronize.
If the associated phase transition is of first order
then the hadronization should proceed through a spinodal phase separation.
We explore here the possibility of identifying the associated clumping
by analysis of suitable $N$-particle momentum correlations.
\end{abstract}


\maketitle

\section{Introduction}

The advent of high-energy nuclear collisions has made it possible 
to create extended quark-gluon plasmas in the laboratory,
thus presenting the prospect of  experimentally probing 
the properties of this fundamental but elusive phase.
Of particular interest is the character of the phase transition
between the plasma and the resonance gas into which it hadronizes.
If the transition is of first order then one would expect 
the hadronization of the plasma to proceed
by means of a spinodal phase decomposition.
The occurrence of this phenomenon might then be utilized 
to signal the phase transition,
as has been successfully done for the liquid-gas phase transition
in nuclear collisions at intermediate energies \cite{ChomazPhysRep389}.

Lattice-gauge calculations suggest
that the character and strength of the phase transformation
depends on the degree of net baryon density
(as governed by the chemical potential $\mu$) 
\cite{KarschNPB605,FodorJHEP0203,Fodor,Ejiri}.
When the system is baryon free, $\mu=0$,
then there is strictly speaking no phase transition at all
but the system evolves from a hadron resonance gas to a plasma
by a continuous crossover as a function of temperature.
As $\mu$ is increased, this crossover grows increasingly abrupt
and becomes a second-order transition 
at a certain critical value $\mu_0$,
above which the transition is truly of first order.
Though it is currently believed that the plasmas 
produced in the midrapidity region at the top RHIC energy 
are too baryon poor to support a first-order transition,
the lattice calculations are not yet sufficiently accurate
to predict the value of $\mu_0$ with confidence.
Thus the phase structure remains open to experimental inquiry
and the present paper explores correlation observables
that may be particularly suitable for such investigations.

A recent schematic analysis of the spinodal growth rates 
in strongly interacting matter \cite{RandrupPRL92}
suggested that the relevant time scales may indeed permit 
the development of spinodal patterns in the hadronizing plasma,
a conclusion supported by more refined dynamical studies 
based on fluid dynamics \cite{Harri}.
Unfortunately, more quantitative predictions cannot be readily made,
since no microscopic dynamical model is yet available for
the phase coexistence region.
Nevertheless, 
because of the potential for obtaining interesting novel information,
and the relative simplicity of the task,
it would seem worthwhile to undertake an analysis of the experimental data
for possible evidence of spinodal decomposition.
Therefore, we seek here to identify observables that could be employed 
for the experimental identification of this phenomenon.

\section{The physical picture}

Spinodal decomposition is a general physical phenomenon
occurring in any system having a first-order phase transition
when it has been quenched away from thermodynamic equilibrium
into the mechanically unstable region of phase coexistence.
(Such a quench is usually achieved by a rapid expansion and/or cooling.)
The phenomenon arises from the presence of a convex anomaly in the relevant 
thermodynamic potential for a spatially uniform system \cite{convex}.
In the region of the anomaly,
the uniform system is thermodynamically and mechanically unstable
and small irregularities will therefore grow spontaneously
as the system seeks to separate into its two coexisting phases.
The growth rate,
which can be calculated by linear-response analysis of the collective modes,
depends on the spatial scale of the disturbance
and the decomposing system will therefore tend to become increasingly orderly
as patterns emerge with a certain characteristic size
corresponding the most rapidly growing instabilities.
It is this transient clumpiness that presents a key
to the search for the phase transition.

The spinodal phenomenon is familiar in many areas of science and technology,
including colloidal aggregation, polymer blends,
binary fluid mixtures and metallic alloys, inorganic glasses,
colloidal aggregation and sticky emulsions.
Technological applications include
spinodal hardening of copper-nickel-tin alloys at the mill
and the production of nanostructured magnetic media
having increased storage density.

Spinodal decomposition also occurs in nuclear collisions
at energies of several tens of MeV per nucleon,
where the nuclear liquid-gas phase transition 
is relevant \cite{ChomazPhysRep389}.
Here the mechanism causes suitably prepared compound systems to
multifragment into several intermediate-mass products of nearly equal size,
in addition to a multitude of nucleons and light fragments.
In these collisions,
the initial compression causes the system to expand sufficiently
for the corresponding temperature-density phase point
to move inside the region of spinodal instability.
The inevitable density irregularities are then amplified
at rates governed by the dispersion relation
for the unstable normal modes associated with the expanded configuration.
Consequently, the most rapidly amplified modes will grow dominant
and the resulting breakup pattern will exhibit a corresponding regularity,
with the prefragments in a particular event 
having approximately equal sizes.
Although much of this regularity is destroyed by the subsequent dynamics,
a sufficient fraction of primordial partitions survive to permit
a clear experimental identification of the signal 
\cite{BorderiePRL86,TabacaruEPJA18}.

As was pointed out in Ref.\ \cite{RandrupPRL92},
an analogous process may happen in high-energy collisions
where a highly excited but rapidly expanding quark-gluon plasma
is produced early on in the collision.
If the phase transition is of first order,
the free energy density exhibits a convex region
into which the bulk of the system is being forced
by the continual expansion.
In this anomalous region, the system has a thermodynamic preference
for reorganizing itself into spatially separate hadron and plasma phases.
One should then expect that the system seeks to acquire a clumpy appearance
with blobs of plasma embedded in a hadronic gas.
The spinodal mechanism tends to make plasma blobs
that have a certain characteristic size.
Ideally these would be positioned fairly regularly 
and the overall collective expansion of the system,
both longitudinally and transversally,
would then cause the overall motions of these plasma blobs 
to exhibit a corresponding regularity in rapidity space.

However, it is important to recognize a fundamental difference 
between the scenarios at low and high energy:
While the liquid blobs produced in multifragmentation
survive as detectable nuclei,
the plasma blobs are only transient and ultimately hadronize,
leaving only hadrons in the final state.
This feature makes it inherently harder to devise
suitable experimental signals of the transient clumping.
In the present paper, we suggest that certain $N$-body momentum correlations
may be suitable for signaling the occurrence of such clumping and
we examine these candidate observables for a number of schematic scenarios
of the type expected if spinodal decomposition occurs.

\section{Longitudinal rapidity}

In our exploratory studies,
we generally represent the clumpy system as a number of distinct sources
which each represents an individual plasma blob
and is assumed to hadronize according to a thermal distribution.
As explained above, the collective expansion causes the spatial configuration
of the blobs to become reflected in their motions,
so we need only specify the collective motions of these sources
(together with their sizes as measured by the number of hadrons emitted).

In order to bring out the key features of the associated correlations,
we first consider only the longitudinal rapidity of the particles
which to a good approximation can be treated analytically.

\begin{figure}[b]
\includegraphics[angle=-90,width=3.6in]{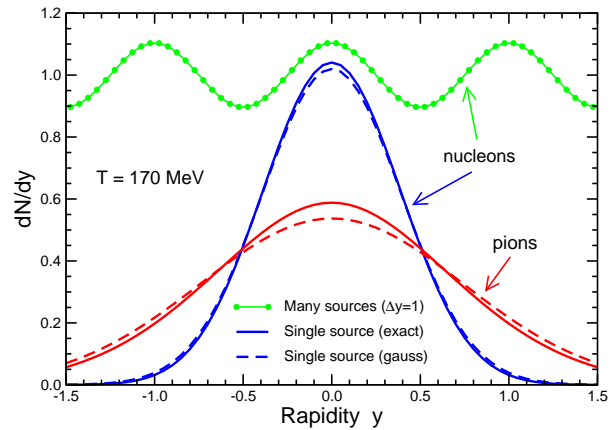}
\caption{\label{f:NN}
The rapidity density resulting from a single thermal source
for either nucleons or pions at $T=170~\MeV$,
together with the nucleon density resulting from a string of sources
with regular rapidity spacings of $\Delta\y=1$.
}\end{figure}

In the rest frame of an individual plasma blob,
the invariant spectrum of the produced hadrons
reflects their thermal weight,
$\rho_1(\p)\equiv Ed^3\nu/d^3\p=d^3\nu/d^2\p_\perp d\y\sim\exp(-E/T)$,
where $E^2=m^2+p_\perp^2+p_z^2=m_\perp^2+p_z^2$ and $dp_z=Ed\y$,
with $m$ being the rest mass of the particular hadron specie considered.
The corresponding rapidity density $\rho_1(\y)\equiv d\nu/d\y$
is obtained by integrating the 3D density $\rho_1(\p)$ 
over the transverse momentum $\p_\perp$ 
and it is to a good approximation of gaussian form,
\begin{equation}\label{G}
\rho_1(\y)\  \equiv\  \int d^2\p_\perp {d^3\nu\over d^2\p_\perp d\y}\
\approx\ {\nu\over\sqrt{2\pi}\sigma_T}\,{\rm e}^{-\y^2/2\sigma_T^2}\ ,\
\end{equation}
where the variance is $\sigma_T^2 = T/(m+T)$
and $\nu$ is the total number of such particles emitted by that source.
At a temperature of $T=170~\MeV$, 
which is near the critical temperature at which the hadronization occurs,
the thermal dispersion of the rapidity distribution 
amounts to $\sigma_T = 0.69, 0.48, 0.38, 0.35$
for $\pi$, $K$, $N$, $\Lambda$, respectively.
The corresponding values at $T=130~\MeV$,
which is closer to the kinetic freeze-out temperature,
are about ten per cent smaller: 0.65, 0.44, 0.34, 0.32.
The thermal rapidity densities are illustrated in Fig.\ \ref{f:NN}
for nucleons and pions at $T=170~\MeV$.

If a collection $\{n\}$ of such sources 
move with the rapidities $\{\bar{\y}_n\}$ and
hadronize independently, the combined one-particle rapidity density would be
\begin{equation}\label{rho1}
\rho_1^{\{n\}}(\y) =
\sum_n \rho_1^{(n)}(\y) \approx\ \sum_n {\nu_n\over\sqrt{2\pi}\sigma_T}\,
{\rm e}^{-{1\over2}(\y-\bar{\y}_n)^2/\sigma_T^2}\ ,
\end{equation}
where  the source $n$ emits $\nu_n=\int\! d\y\,\rho_1^{(n)}(\y)$ particles.
In Fig.\ \ref{f:NN}, the resulting rapidity density is illustrated 
for nucleons in the simple situation where the sources are similar
and have a constant rapidity separation of $\Delta\y=1$.
If $\Delta\y$ is decreased below unity,
the combined nucleon density rapidly approaches a constant,
while the corresponding pion density is practically flat 
already for $\Delta\y=1$.
This schematic illustration suggests that it is preferable
to focus on hadrons that are heavy
in order to reduce the effect of the thermal smearing.
We shall therefore carry out our analyses for protons.
(Since the determining quantity is the mass of the considered hadron,
the term ``proton'' here includes also $\bar{p}$.)
From the above numbers one may judge the effect 
of considering other hadronic species.

The two-particle rapidity density  
is obtained by performing a corresponding transverse projection of
the 3D two-particle density $\rho_2(\p_1,\p_2)\equiv d^6\nu/d^3\p_1d^3\p_2$.
For a collection of sources $\{n\}$ it is generally of the form
\begin{equation}
\rho_2^{\{n\}}(\y_1,\y_2) = 
\sum_n \rho_2^{(n)}(\y_1,\y_2) +
\sum_{n\neq n'} \rho_1^{(n)}(\y_1)\rho_1^{(n')}(\y_2)\ ,
\end{equation}
where $\rho_2^{(n)}(\y_1,\y_2)$ accounts for the situation
when both particles arise from the same source $n$. 
This term is here normalized to the total number of pairs
emitted by that source,
$\int d\y_1\int d\y_2\, \rho_2^{(n)}(\y_1,\y_2)=\nu_n(\nu_n-1)$,
thus ensuring that
$\int d\y_1 \int d\y_2\, \rho_2^{\{n\}}(\y_1,\y_2)=\nu(\nu-1)$
where $\nu\equiv\sum_n \nu_n$ is the total number of particles emitted
from the particular source configuration denoted by $\{n\}$.
(A division of $\rho_2$ by $2!$ should be made 
when the two observed particles are indistinguishable, 
but this is immaterial here.)

It is especially instructive to consider the distribution 
of the {\em rapidity difference} $\y_{12}(\y_1,\y_2)\equiv\y_1-\y_2$,
\begin{eqnarray}\nonumber
F^{\{n\}}_{\rm 1D}(\y) &\equiv& \int\!d\y_1\!\int\!d\y_2\, 
\delta(\y-\y_{12}(\y_1,\y_2))\, \rho^{\{n\}}_2(\y_1,\y_2)\\
&=& \sum_n {\nu_n(\nu_n-1)\over 2\sqrt{\pi}\sigma_T}\,
{\rm e}^{-\y^2/4\sigma_T^2} \\ \nonumber
&+& \sum_{n\neq n'} {\nu_n \nu_{n'}\over 2\sqrt{\pi}\sigma_T}\,
{\rm e}^{-(\y-\bar{\y}_n+\bar{\y}_{n'})^2/4\sigma_T^2}\ ,
\end{eqnarray}
which depends only on the absolute value of the rapidity difference, 
$|\y_{12}|$,
and is normalized to the total number of pairs emitted,
$\int d\y\,F^{\{n\}}_{\rm 1D}(\y)=\nu(\nu-1)$.

In order to incorporate some degree of irregularity 
in the source configurations,
we assume that the number of particles emitted from a given source, $\nu_n$,
is a random variable (usually a Poisson distribution)
and that the source rapidity $\bar{\y}_n$ has a normal distribution around
$\bar{\bar{\y}}_n$ with the (common) dispersion $\sigma_\parallel$.
Making an ensemble average $\prec\cdot\succ$ over such sources,
we then obtain the probability distribution for the rapidity difference,
which is normalized to unity, $\int\!d\y\,P_{\rm 1D}(\y)=1$,
\begin{eqnarray}\nonumber
&~& \prec F^{\{n\}}_{\rm 1D}(\y)\succ\ =\
\prec \nu(\nu-1)\succ P_{\rm 1D}(\y)\\
&=& \sum_n {\prec\nu_n(\nu_n-1)\succ\over [4\pi\sigma_T^2]^{1\over2}}\,
{\rm e}^{-\y^2/4\sigma_T^2} \\ \nonumber
&+& \sum_{n\neq n'} 
{\prec\nu_n\succ\prec\nu_{n'}\succ
\over[4\pi(\sigma_T^2+\sigma_\parallel^2)]^{1\over2}}\,
{\rm e}^{-(\y-\bar{\bar{\y}}_n+\bar{\bar{\y}}_{n'})^2/
4(\sigma_T^2+\sigma_\parallel^2)}\ .
\end{eqnarray}
The effect of the random selection of the source rapidities $\{\y_n\}$
is to broaden the individual distributions,
in effect adding $\sigma_\parallel^2$ 
to the variance $\sigma_T^2$ in (\ref{rho1}).
This in turn makes the mixed-source contributions to $P_{\rm 1D}(|\y_{12}|)$
correspondingly broader,
while the same-source contributions  remain unaffected.
As a result, the same-source terms stand out more prominently.

To present the results on a robust and instructive form,
we divide the correlated probability distribution
by the corresponding uncorrelated result.
While this latter quantity can generally be obtained by the event mixing
(drawing the two particles from different events),
the boost invariance of the present event ensemble implies that 
it is given simply in terms of the average rapidity density,
$P_{\rm 1D}^0(\y_{12}) = (1/\bar{\nu})d\bar{\nu}/d\y$.
We thus consider the {\em reduced 1D rapidity correlation function},
\begin{equation}
C_{\rm 1D}(|\y_{12}|)\ \equiv\ 
{P_{\rm 1D}(\y_{12}) + P_{\rm 1D}(-\y_{12})
 \over P_{\rm 1D}^0(\y_{12}) + P_{\rm 1D}^0(-\y_{12})}\ -\ 1\ .
\end{equation}
Apart from finite-number effects,
this quantity does not depend on the magnitude of $d\bar{\nu}/d\y$
and the subtraction of unity guarantees that
it approaches zero for large values of $|\y_{12}|$,
where the particles are uncorrelated.

\begin{figure}[b]
\includegraphics[angle=-90,width=3.6in]{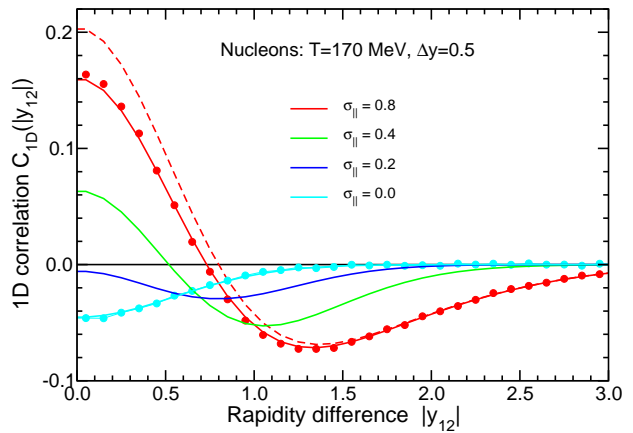}
\caption{\label{f:C}
The two-proton correlation $C_{\rm 1D}=P_{\rm 1D}/P_{\rm 1D}^0-1$
as a function of the difference in longitudinal rapidity $|\y_{12}|$,
for similar sources having an average spacing of $\Delta\y=0.5$
with individual rapidity dispersions of $\sigma_\parallel=0.0,0.2,0.4,0.8$.
The curves shown are calculated with the gaussian approximation, 
which is analytical, but the numerical sampling of the exact distribution
yields very similar results,
as illustrated by the solid circles for $\sigma_\parallel=0.0, 0.8$.
If the actual multiplicities $\{\nu_n\}$ have a Poisson distribution
then the curves move up:
The curve for $\sigma_\parallel=0.0$ becomes zero to a good approximation,
while the result for $\sigma_\parallel=0.8$ is shown by the dashed curve.
}\end{figure}

Figure \ref{f:C} shows the resulting correlation
 for a number of schematic scenarios
that all use an average rapidity spacing of $\Delta\y$=0.5 but have
various degrees of rapidity smearing as specified by $\sigma_\parallel$.
When $\sigma_\parallel$ vanishes 
the sources have fixed equidistant separations
and the distribution $F_{\rm 1D}(\y_{12})$ is then periodic.
Nevertheless, the correlation function $C_{\rm 1D}(|\y_{12}|)$
exhibits a depression near $|\y_{12}|\approx0$,
since for each particular source 
there is only one source with the same rapidity
while there are two with rapidities differing by $\pm\Delta\y$,
$\pm2\Delta\y$, {\em etc}.
Thus, curiously, a high degree of regularity in the source configuration
will be reflected in a deficit of particle pairs 
that have very similar rapidities.
By contrast, when the individual source rapidities are endowed
with a random deviation from regularity,
the contributions from pairs emitted by different sources
will be smoother, as noted above,
making the same-source contribution stand out more clearly.
As $\sigma_\parallel$ is increased, 
the same-source term grows increasingly prominent
and it approaches the form that would result from a single thermal source.
The figure also illustrates the good quality of the analytical
(gaussian) approximation (\ref{G}).

These results have been obtained for sources that all contribute
the same number of particles $\nu_n$.
In order to illustrate the sensitivity to
random variations in the multiplicities $\{\nu_n\}$,
the figure also shows results obtained
by using a Poisson distribution for $\nu_n$.
This additional source of irregularity generally increases the correlation.

The schematic analysis described above suggests that
the two-particle rapidity correlations may be sensitive to 
the degree of clumping in the emitting system.
However, some detectors ({\em e.g.}\ the STAR TPC)
have rather limited rapidity coverage
while the range of acceptance tends to be larger
in the transverse direction.
It may therefore be of interest to perform analyses
that invoke the full three-dimensional particle motion.

\section{Relative rapidity in 3D}

As a natural generalization to three dimensions,
we shall now consider the (Lorentz invariant) {\em relative rapidity},
{\em i.e.}\ the rapidity of one particle as seen from the other,
\begin{equation}\label{y12}
\y_{12}(p_1,p_2)\ =\ 
\ln[\gamma_{12}+\sqrt{\gamma_{12}^2-1}]\ \geq\ 0\ ,
\end{equation}
where $\gamma_{12}$ is the Lorentz factor for the relative motion,
$m_1m_2\gamma_{12} = p_1\cdot p_2 = E_1E_2-\p_1\cdot\p_2$.
The analysis procedure is then the same as above.
For a given sample of events,
we extract the (normalized) probability distribution 
for the relative rapidity $\y_{12}$
from the correlated two-particle density $\rho_2(\p_1,\p_2)$,
\begin{equation}
P_{\rm 3D}(\y)\ \equiv\	
\prec\delta(\y_{12}-\y)\ \rho_2(\p_1,\p_2)\succ
/\prec \rho_2(\p_1,\p_2)\succ\ .
\end{equation}
Subsequently we divide it by the uncorrelated reference density 
$P_{\rm 3D}^0(\y)$, which can be obtained by event mixing,
and we finally subtract unity to obtain the reduced  correlation function
for the relative rapidity,
\begin{equation}
C_{\rm 3D}(\y)\ \equiv\ 
{P_{\rm 3D}(\y) \over P_{\rm 3D}^0(\y)}\ -\ 1\ .
\end{equation}
We note that even the uncorrelated function $P_{\rm 3D}^0(\y)$
has a non-trivial form reflecting the composite nature of the emitting system.
In the non-relativistic limit, 
where $\y_{12}$ is the relative speed $v_{12}=|\bfv_1-\bfv_2|$, 
a single thermal source having $d^3\nu/d^3\p\sim\exp(-p^2/2mT)$ yields
$P_{\rm 3D}(v)\sim v^2\exp(-mv^2/4T)$.

We examine this observable for schematic scenarios
that incorporate some degree of transverse flow
in addition to the overall longitudinal expansion.
For this purpose, we generate an ensemble of source configurations
by making random variations relative to a suitable scaffold configuration 
of individual thermal sources in rapidity space.
The scaffold configuration consists of sources situated 
at the vertices of $N_\phi$-sided equilateral polygons ({\em e.g.}\ squares)
that are oriented perpendicular to the $z$ axis 
and placed with regular spacings $\Delta\y$ in longitudinal rapidity.
(The (mean) azimuthal separation between two neighboring scaffold sources 
in a given polygon is thus $\Delta\phi=2\pi/N_\phi$ and each polygon 
is rotated by half that amount relative to its neighbors.)
On the average, each individual thermal source 
emits the same number of particles $\bar\nu_n$, 
and the magnitude of its transverse flow rapidity is $\y_T$.
The mean rapidity density is thus
$d\bar{\nu}/d\y=N_\phi\bar{\nu}_n/\Delta\y$.

\begin{figure}[h]
\includegraphics[angle=-90,width=3.6in]{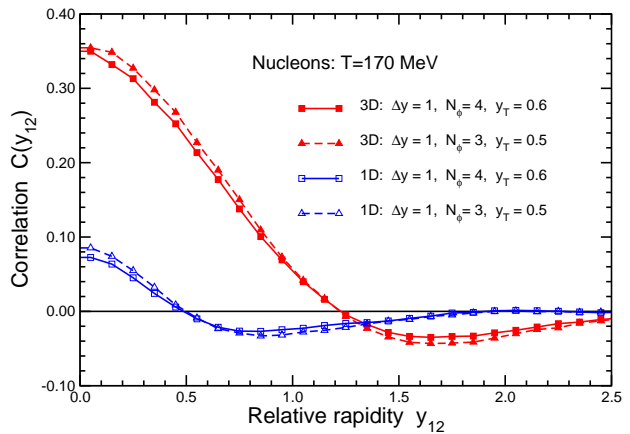}
\caption{\label{f:3}
The reduced 1D and 3D rapidity correlation functions
for our standard scenario having $N_\phi=4$ (see text)
or a modified scenario using $N_\phi=3$ with the transverse flow
reduced to $\y_T=0.5$ in order to approximately preserve
the recession speed between neighboring sources.
}\end{figure}

Relative to the scaffold configuration,
the actual velocity of each source $n$ is obtained
by adding a random deviation
with regard to both its longitudinal rapidity 
and its transverse flow vector,
as well as to the number of particles emitted.

For illustrative purposes, let us consider a scenario
where the polygons are squares ({\em i.e.}\ $N_\phi=4$)
that are placed with rapidity separations of $\Delta\y=1$
and whose corners are endowed with a transverse flow rapidity of $\y_T=0.6$.
It then follows that the relative rapidity between
two neighboring sources in the same polygon is approximately the same 
as that between neighboring sources in adjacent polygons
(namely $\approx$1.20).
Relative to this scaffold configuration,
we assume that the actual number of protons emitted by a given source
is governed by the corresponding Poisson distribution,
and we take the dispersion of the longitudinal rapidity of a given source,
$\sigma_\parallel$, as well as the dispersion of its transverse flow rapidity,
$\sigma_\perp$, to be 0.3.

The resulting 1D and 3D rapidity correlation functions
$C_{1D}(|\y_{12}|)$ and $C_{3D}(\y_{12})$ are shown in Fig.\ \ref{f:3}.
They are qualitatively similar:
their peak at $\y_{12}\approx0$ is followed by a shallow dip 
with an eventual slow rise towards the limiting value of zero.
(The results for $C_{\rm 1D}(\y)$ agree well with 
what would be obtained with the 
one-dimensional analytic approximation employed above,
so if one were interested in only the longitudinal rapidities
there would be no need to carry out the entire 3D simulation.)
It is evident that the 3D correlation function
exhibits a significantly stronger signal,
which is a result of the additional outwards motion of the sources.
It is thus advantageous to invoke the entire three-dimensional particle motions
in the correlation analysis.

The resulting correlation functions are (nearly) independent
of the overall multiplicity,
as specified by the average rapidity density $d\bar{\nu}/d\y$.
Furthermore, their dependence on the specific geometrical arrangement
of the sources in flow space is primarily through the average recesssion
speed between neighboring sources, as illustrated in Fig.\ \ref{f:3}.
In addition, as we have discussed above,
the correlation signal generally increases with
the degree of fluctuation in the source sizes and their motion.
Thus the observable is sensitive to the dynamical features 
of the clumping in the system.

\section{\bold{N}-body correlations}

Since a variety of dynamical processes produce two-body correlations, 
it may be advantageous to consider also correlations 
between more than two particles,
to thus be able to better discriminate between the mechanisms.
We therefore consider the analysis of the $N$-body density
$\rho_N\{\p_n\}=E_1\cdots E_Nd^{3N}\nu/d^3\p_1\cdots d^3\p_N$.

A particularly convenient and instructive correlation observable
is the internal kinetic energy per particle,
\begin{equation}
\kappa_N\{\p_n\}\ \equiv\ {1\over N}\left[
\left[P\{\p_n\}\cdot P\{\p_n\}\right]^{1\over2} - \sum_{n=1}^Nm_n\right]\ ,
\end{equation}
where $P\{\p_n\}=\sum_np_n=\sum_n(E_n,\p_n)$
is the total four-momentum of the $N$ observed particles.
The normalized probability distribution of this quantity is then
\begin{eqnarray}\nonumber
P_N(\kappa) &=& \prec\prod_{n=1}^N\left[\int d^3\p_n\right]
\delta(\kappa_N\{\p_n\}-\kappa)\ \rho_N\{\p_n\}\succ\\
&/& \prec\prod_{n=1}^N\left[\int d^3\p_n\right]\rho_N\{\p_n\}\succ\ ,
\end{eqnarray}
For a single thermal source in the non-relativistic limit, $T\ll m$, 
this probability distribution is given analytically,
\begin{equation}
P_N(\kappa)\ =\ {1\over\kappa}\, {1\over\Gamma({3\over2}N-{3\over2})}
\left({N\kappa\over T}\right)^{{3\over2}N-{3\over2}} {\rm e}^{-N\kappa/T}\ .
\end{equation}
Thus $P_N(\kappa)\sim \kappa^{(3N-5)/2}\exp(-\kappa/T)$
which peaks at the value 
$\kappa=(\mbox{${3\over2}$}-\mbox{${5\over2N}$})T$.

The motivation for considering this observable is the expectation
that when the $N$-body momentum distribution is clumped,
the sampling of $\kappa_N$ will yield an enhancement around 
the typical ({\em i.e.}\ thermal) kinetic energy in the individual source,
relative to what would occur for a structureless distribution.
In order to bring out this signal, we proceed as before 
and compare the correlated distribution $P_N(\kappa)$,
obtained by sampling all the $N$ particles from the same event,
with the corresponding uncorrelated distribution $P_N^0(\kappa)$
obtained by sampling the particles from $N$ different events.
This yields the reduced correlation function for the internal kinetic energy,
\begin{equation}
C_N(\kappa)\ \equiv\ {P_N(\kappa) \over P_N^0(\kappa)}\ -\ 1\ .
\end{equation}

Contact with the two-body rapidity correlations considered earlier can be made
by noting that the internal kinetic energy per particle
determines the characteristic internal rapidity,
$\y_N=\ln[\gamma_N+\sqrt{\gamma_N^2-1})$,
where $\gamma_N=\kappa_N/m+1$ (assuming that all $N$ masses are equal).
In particular, for $N=2$ the internal rapidity is simply half 
of the relative rapidity $\y_{12}$ introduced in Eq.\ (\ref{y12}).

\begin{figure}[t]
\includegraphics[angle=-90,width=3.6in]{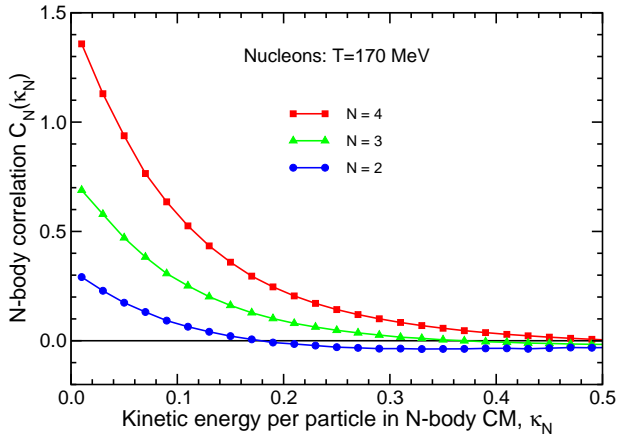}
\caption{\label{f:4}
The reduced $N$-body correlation function $C_N(\kappa_N)$ for $N=2,3,4$
for the internal kinetic energy per particle in $N$-body systems $\kappa_N$
in the same scenario as used for Fig.\ \ref{f:3}.
The corresponding results
in the presence of a strong (1:2) elliptic flow, 
using $\y_x=0.8$ and $\y_y=0.4$ rather than $\y_x=\y_y=0.6$,
are also shown (dashed curves).
}\end{figure}

Illustrative results for $C_N(\kappa)$
are shown in Fig.\ \ref{f:4} for $N=2,3,4$.
The correlation signal grows more prominent 
as the correlation order $N$ is increased
(its strength approximately doubles for each particle added),
as expected because it becomes increasingly unlikely that $N$ momenta
sampled from a structureless distribution would all be nearly similar.
This feature is the reason why higher-order correlations may be advantageous.
On the other hand, since the signal receives its support
from a relatively small region of the $N$-body phase space, 
the required counting statistics increases in a factorial fashion with $N$, 
thus presenting a practical limit to the order
of correlation that can be addressed with a given set of data.
However, the information conveyed by the higher-order correlations
is progressively more effective as a discriminator
between various possible underlying dynamical mechanisms.

It is also noteworthy that $C_n(\kappa)$ falls off approximately as 
$\sim\exp(\kappa/T_N)$ with $T_N\approx (1-N^{-1})$,
so once the correlation has been identified its shape can,
in principle, be utilized to infer the temperature of the emitting sources.

In the above scenarios, we have for simplicity assumed that 
the collective transverse flow is axially symmetric.
However, a significant degree of azimuthal anisotropy has been observed
\cite{ReisdorfARNPS47,AckermannPRL86,AdlerPRL87,AdcoxPRL89,AltPRC68}
and we therefore also consider scenarios containing elliptic flow,
which is the most prominent type.
For this purpose, we give the mean transverse source rapidity $\y_T$
an elliptic azimuthal variation characterized by its values 
in the $x$ and $y$ directions.
(In the analysis of experimental data, this would correspond to performing 
an azimuthal rotation on each event to line up the flow axes.)
By this means, 
the flow-induced $N$-body correlations are included also in the 
mixed-event distribution by which the same-event distribution is divided
and, as a consequence, they are largely eliminated.
For example, when this procedure is employed
even a scenario having a strong elliptic flow,
obtained by using $\y_x=0.8$ and $\y_y=0.4$ rather than $\y_x=\y_y=0.6$
(so $\y_x:\y_y=2:1$),
produces no noticeable difference in the resulting 
curves for $C_N(\kappa)$ shown in Fig.\ \ref{f:4}.

\section{Discussion}

Spinodal decomposition, if indeed it occurs,
has considerable potential utility as a probe
of the hadronization phase transition.
It would therefore seem worthwhile to undertake a systematic
analysis the data for the purpose of identifying its occurrence.
For this task, the specific observables examined above may be useful,
either for direct use in the analysis
or as a stimulation for the development of alternate observables
that may provide signals of the spinodal phase separation.

The present exploratory study has ignored several complicating features.
Perhaps most important is the assumption that the individual hadrons
propagate directly to the detectors without further interaction
after their initial formation.
In reality there may well be a considerable amount of rescattering 
among the hadrons after their formation.
Fortunately, 
such rescattering does not necessarily destroy the spinodal pattern,
since the particles would interact primarily with others
originating from the same source.
In fact, if this were strictly true,
then the rescattering would merely reduce the kinetic freeze-out temperature
which in turn would reduce the thermal smearing and thus
in effect help to make the source structure stand out more clearly
(though this effect is, as we have noted, relatively modest).
This important aspect could be explored quantitatively
within microscopic simulation models (such as RQMD),
in which one may specify an initial lumpy distribution of hadrons
and then propagate the system until the collisions cease.
This would obviously be an interesting undertaking
that could help to clarify the utility of the correlation function
as a tool for signaling clumping.

We have here primarily been concerned with ascertaining
to what degree existing clumping can in fact be revealed
through suitable correlation observables.
But before embarking on such an analysis of the data,
one should recognize that a variety of mechanisms 
may produce $N$-body enhancements 
and the analysis therefore needs to be made with appropriate care.
Fortunately, as we have demonstrated for the case of elliptic flow,
the complications from directed collective flow can be largely eliminated,
insofar as the flow is well understood.

Another possible complication arises from the production of hadronic jets,
which contribute strong $N$-body correlations.
However, while the momenta of the jet particles 
are approximately unidirectional, they are not clumped.
Furthermore, the relative importance of jet production increases with the
beam energy, as does the mean energy of the jet particles,
while spinodal decomposition most likely occurs only at relatively low
beam energies (where the chemical potential is sufficiently large)
and leads to hadrons with thermal energies.
Therefore jets may not present a serious obstacle.

As noted in the beginning,
the matter produced in the central rapidity region at RHIC
is expected to have a subcritical baryon density, $\mu<\mu_0$,
so the transformation from plasma to hadron gas occurs smoothly.
Therefore, by analyzing data sets corresponding to
scenarios having different degrees of baryon content,
it should be possible to explore physical scenarios 
ranging from subcritical, where no phase separation occurs,
to supercritical, where spinodal decomposition is favored.
Such a scan with respect to $\mu$ could conceivably be made at RHIC
by moving from midrapidity towards the fragmentation regions
where the baryon fraction is generally higher.
Alternatively one might invoke also data from the SPS regime or,
in the future, from the new facility being built at GSI,
where the net baryon contents is significantly higher.

Obviously, the observation of spinodal clumping would reveal 
a novel phenomenon whose existence has a direct bearing on character of the
phase structure of strongly interacting matter
and which could then be utilized as a means for 
obtaining quantitative information about the equation of state.
Conversely, if such experimental investigations were to suggest
the general absence of spinodal decomposition,
it would imply that either the phase transition is never of first order or,
even where a first-order transition is present,
the collision dynamics fails to bring the system into the plasma phase
or the spinodal growth is too slow 
for the decomposition to develop sufficiently.
In any case, the experimental information would establish 
useful constraints on the basic models.
It therefore seems worthwhile to consider undertaking such an analysis effort
and we hope that this study will provide some encouragement towards this.

~

Helpful discussions with Fabrice Reti\'ere are acknowledged.
This work was supported by the Office of Energy Research,
Office of High Energy and Nuclear Physics,
Nuclear Physics Division of the U.S. Department of Energy
under Contract No.\ DE-AC03-76SF00098.



                        \end{document}